# GridEmail: A Case for Economically Regulated Internet-based Interpersonal Communications

Manjuka Soysa[1], Rajkumar Buyya[1], and Baikunth Nath[2]

[1]Grid Computing and Distributed Systems Laboratory
[2]Department of Computer Science and Software Engineering
University of Melbourne, Australia
Email: {manjuka, raj, bnath}@cs.mu.oz.au

## Abstract

*Email has emerged as a dominant form of electronic communication between people. Spam is a major problem for email users, with estimates of up to 56% of email falling into that category [13]. Control of Spam is being attempted with technical [38-44] and legislative [19 -36] methods. In this paper we look at email and spam from a supply-demand perspective. We propose Gridemail, an email system based on an economy of communicating parties, where participants' motivations are represented as pricing policies and profiles. This system is expected to help people regulate their personal communications to suit their conditions, and help in removing unwanted messages.*

## 1. Introduction

Email was designed for communication in trusted environments, and later found application as a ubiquitous form of communication. The protocols and architecture of email are designed for a simple interaction to send a message from one person to another. Spam and other forms of abuse have taken advantage of the lack of security features in email. According to statistics maintained by Brightmail, spam accounted for about 56% of emails in November 2003, up from 18% in 2002 [13]. The Nucleus Research has reported that in 2003 spam costs the average organization $874 per employee per year [51]. Worldwide cost of spam has been estimated to be $113billion [52] and it is rapidly increasing.

The Anti-Spam Research Group (ASRG) of the Internet Research Task Force was formed to survey, evaluate and recommend solutions to the growing problem of spam. The ASRG has created a list of weaknesses in current email systems [1], and produced a taxonomy of current solutions to the spam [2], shown in Figure1.

The continuing problems caused by spam emails indicate that the current techniques are not providing a completely satisfactory solution. It has been recognized that effective solutions may emerge from major changes to the Simple Mail Transfer Protocol (SMTP) [3]. There have been several proposals to the Internet Engineering Task Force (IETF) for such major changes. The Adaptive Mail Delivery Protocol (AMDP) proposal [10] aims to cut spam at the source. The Trusted Email Open Standard (TEOS) proposal from the E-privacy group [11] aims to add extra security and trust features.

Anti-Spam legislation has been introduced in the USA, various European countries and Australia [19-36]. Their effectiveness on reducing spam remains to be seen [37].

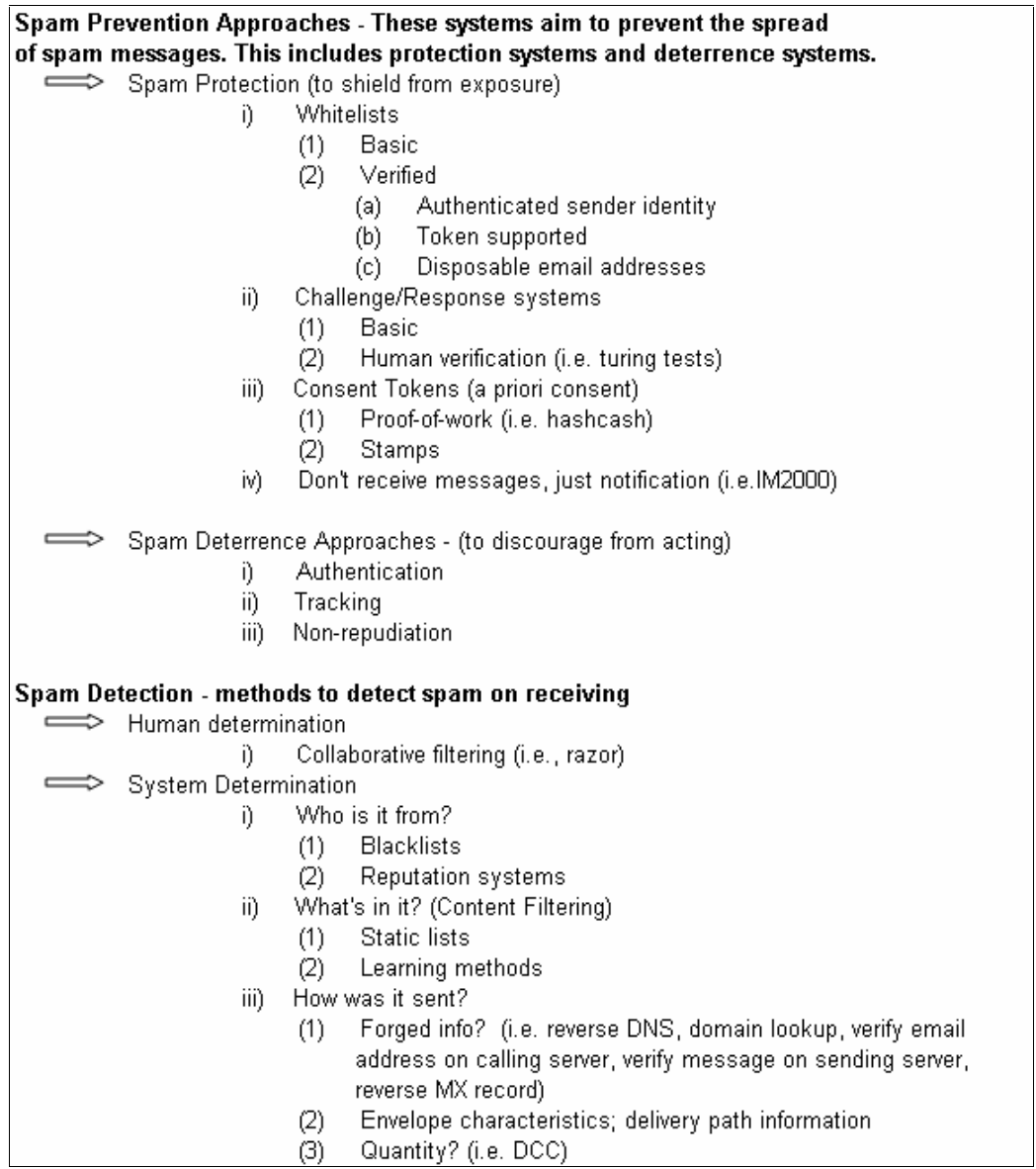

**Figure 1 – Taxonomy of spam-fighting techniques (Ref [2]).**

## 2. The economics of spam

The economics of spam have been studied by several parties [7, 8, 14]. These studies show that spammers have to send millions of emails to get a sufficient return. Therefore spam is viable only due to the lack of a significant per-recipient cost for sending email. Micropayments [15], CPU cycles [9] and traffic shaping [12] have been suggested as methods for incurring a cost to the sender to discourage sending emails to a large number of recipients. CPU-cycle based systems have been used in research environments, while commercial systems are available for micropayments and traffic shaping.

The Centipaid micropayment system [15] can work as a filter that quarantines messages that do not have a payment attached, and whose senders are not in a white list. Senders of quarantined messages get a reply directing them to a web page where payments can be attached.

At its core, email provides a method of exchanging information, and email communications can be viewed as economic transactions that are subject to supply and demand behavior [14]. While the systems described above introduce an economic disincentive for sending spam, they do not attempt to provide ways to optimize the economic benefit to the sender or the recipient. This is what Gridemail aims to provide.

## 3. Rationale of Gridemail

People are motivated to send and receive messages for different reasons. Both sending and receiving parties consume resources in the exchange of information, and would therefore expect to gain something. It maybe a tangible benefit as in the case of a sale resulting from an advertisement, or an intangible benefit like goodwill gained from a greeting. With Gridemail, we want to help both parties to maximize the benefit of an exchange.

To pre-empt the possibility of being used by spammers, we want a system that will support significant per-recipient costs. A person wanting to send multiple messages on such a system, will want to maximize the return on the resources spent.

A recipient, who spends valuable time in reading messages, would like to maximize the benefit by reading only relevant or important messages. Current solutions provide minimal control to the recipient to differentiate and choose the messages being received.

When resources such as storage or internet services are placed on a network, measures are put in to control access to the resources. Quality of Service (QoS) and Classes of Service (COS) are used to differentiate users and provide services in a selective way.

Therefore it is only natural to consider the time and mental effort of a recipient as a valuable resource, the use of which has to be regulated. However, neither quality of service nor congestion was a consideration in the design of SMTP, as initially it was used as a medium of communication between trusted parties. Other end-user messaging technologies (fax, SMS) largely rely on a single charge to provide some economic control. Essentially each messaging system provides a single class of service, leaving the recipient open to abuse or overload if the class is not secure or the price is too low.

The Gridemail framework described in this paper allows users to specify differentiated classes of service, each with its own policies. Each class of service represents the users' policy and conditions for processing a certain type of message. The sender has to select a class of service suitable for a given message, and then satisfy the conditions to use it.

Gridemail is not merely a spam fighting technique. It is a system that provides a framework to manage all personal message-based communications. Considering trends towards unified messaging and pervasive computing, different levels of QoS for messaging is likely to be a requirement for future applications. People have an increasing number of communication channels available, each with different qualities. Without a framework like Gridemail, managing and preventing abuse in each channel separately becomes a time consuming task for individuals.

## 4. Related areas of research

The techniques proposed in Gridemail have some similarities to congestion control and customer classes principles. This area has been widely researched since 1969 [5, 16-18], using queuing theory and models, for application in controlling access to limited resources. Economic methods for regulating access to distributed resources are a related area receiving current attention [45].

Adopting such techniques to personal communication systems present some challenges. The behavior of a person providing a service is significantly different from that of a networked machine. Modeling of the proposed system has to take into account these differences.

Access patterns of a network resource can be measured easily using the logs. Usage patterns of e-mail are harder to measure due to privacy considerations, and also tend to vary largely from person to person. Assumptions have to be made about behavior patterns when modeling the system.

## 5. Quality of Service Factors

The major measures of QoS, applied to an end-user messaging service are described below. Some are derived from QoS factors for webservices described in [4].

- Availability: The availability of the user to receive messages at a given time.
- Accessibility: Accessibility of service to different senders.
- Integrity: Level to which integrity of a message data and delivery status are maintained.
- Performance: Latency of delivery and processing (maximum delivery time, etc).
- Reliability: Degree of being capable of maintaining the service and service quality. The number of failures per month or year represents a measure of reliability of a Web service. In another sense, reliability refers to the assured and ordered delivery for messages being sent and received by service requestors and service providers.
- Flexibility: Extent to which support is provided for rules, compliance with standard related to message types, content and formats.
- Security: Security is the quality aspect of the service of providing confidentiality and non-repudiation by authenticating the parties involved, encrypting messages, and providing access control. The service provider can have different approaches and levels of providing security depending on the service requestor.
- Recipient Properties: Properties of the recipient that affect the ability process the message, such as interests, areas of expertise, trust and reputation.

The above QoS factors can be provided in different layers including Network, System, Channel and User layers. Issues related to QoS factors in the Network layer are beyond the scope of this paper. System layer refers to QoS factors that are provided by the Gridemail software. Device QoS refers to inherent properties of the Channel and device used to read messages. User QoS refers to properties and behavior of the user processing the messages.

The configuration task in setting up Gridemail for a recipient involves defining different classes of service and their policies. Each may have different mixes of QoS in the different layers. Gridemail will direct received messages to different channels depending on the class of service chosen by the sender, and the policy configuration of the recipient.

## 6. High-level architecture of Gridemail

The main components of the Gridemail system are illustrated in Figure 2. A person who wants to send a message composes the message and chooses a profile for the message. The message is then transmitted to the Gridemail Sender Service, which is an internet service located at an ISP or corporate network. After negotiating a suitable class of service, the message is transferred to a queue in the Receiver Service, where it is held until sent to the browsing application.

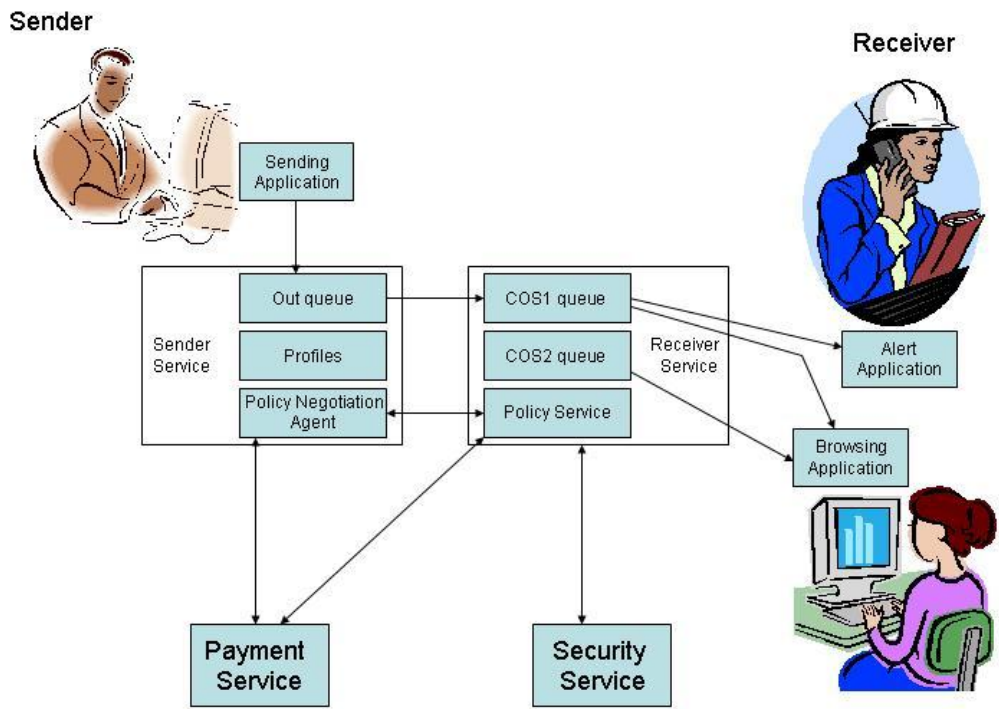

**Figure 2 – High-level architecture of Gridemail.**

**Sending Application**
The Sending Application may execute on a phone, desktop computer, or some other device with Internet connectivity. This application is used to compose the message. Depending on the device, the application may have varying capabilities for creating different types of content. Once the message is composed, the sender has to specify a Recipient and a Profile. The Recipient may be specified by providing the equivalent of an email address. The Profile indicates the QoS requirements of the message. The Sender can select from pre-defined Profiles (that reside in his Messaging Service or Device), use a default Profiles of the Recipient, or create a custom Profile. The profile may contain information related to the sender's valuation of message in terms of parameters such as budget that he/she is willing to invest for posting it.

**Sender Service**
The Sender Service and Receiver Service may be components of a single Messaging Service. They are explained separately because they play different roles when sending or receiving. The Sender Service has an queue that holds a sender's message until it is transferred to the Receiver Service. The Policy Negotiation Agent is responsible for contacting the Receiver Service, negotiating a suitable class of service, security and payment requirements, and then transferring the message. Any problems should be reported back to the Sending Application.

**Receiver Service**
The Receiver Service implements the queues for different classes of service. These queues hold incoming until transferred to the recipient's browsing application. For some classes of service, a alert message may also be sent to inform the arrival of a message. The Policy Service component is responsible for specifying availability and pricing for the QoS requirements specified by the Sender Service.

**Security Service**
Some classes of service will require authentication of the Sender. A Certification Authority may act as a third party to authenticate the Sender.

**Payment Service**
In cases where the Receiver Service specifies a price to be paid for queuing a message, the payment can be made through this third party service.

**Browsing Application**
The Browsing Application is a mail browser running on an internet connected device. This will pull messages from the relevant queue, and allow the recipient to process its content. This application will also act as the storage for offline browsing of messages.

**Alert Application**
Some classes of service will need urgent notification when a message is received. In these cases, an alert may be sent to an Alert Application on a high-availability device such as a mobile phone. The message remains in the queue until it is sent to the Browsing application later.

## 7. Gridemail Operational Model

The activity flow of Gridemail will vary depending on whether person to person, bulk, or automated messaging is taking place. Figure 3 illustrates the activities involved in person to person communication.

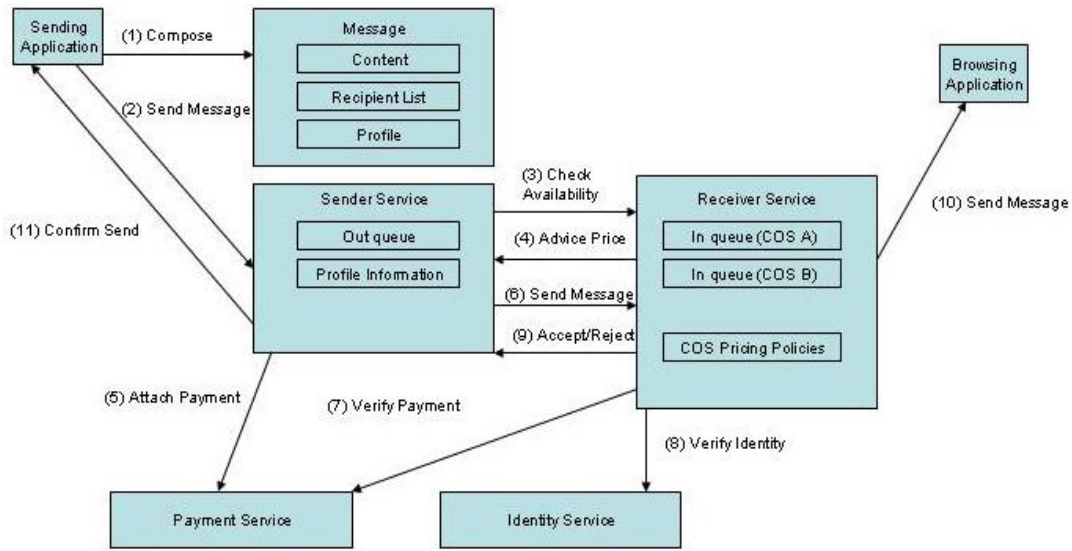

**Figure 3 – Activity flow for using person to person communication.**

The basic steps for a person to send a message are:
1. The sender has to compose a message using the sending application.
2. The sending application transfers the message to his messaging service, and specifies the recipient and a profile for the message.
3. The Sender Service contacts the Receiver service, and looks for a class of service suitable for the message profile.
4. The Receiver Service specifies price and availability for the selected class of service.
5. The sender's service checks whether the price is within the profile's limit, and if so, a payment is attached.
6. The message is sent with the payment.
7. The Receiver Service verifies the payment.
8. The Receiver Service verifies the identity of the sender.
9. Sender's service provides feedback to the sender's device.
10. Depending on the class of service selected, the Receiver Service may push the message to the receiver's device, or keep it till the message is retrieved.
11. A verification or rejection message is sent to the sender's device.

In addition to monetary payments, different scoring functions may be used:
- Score based on source - points if it is from a friend, business partner, etc
- Score based on a reply-paid stamp - points if it contains a stamp which indicates the recipient is expecting the mail
- Score based on format – points if it is in a certain format that helps automated processing (eg: an ICAL message requesting an appointment)

It should also be noted that rejecting messages happens in conventional email only when the Inbox of a recipient is full. It is a physical resource restriction, fundamentally different from a pragmatic decision based on a recipient's ability to process messages. The sender's messaging service has to provide feedback to the sender when the message can not be sent immediately. Automatic re-tries may be an option.

## 8. Modeling Gridemail

Analysis of the Gridemail model is necessary to show how it provides a benefit compared to conventional email. The benefit can be measured from sender, recipient, intermediary, or social viewpoint. The definition of benefit can also vary with the scenario.

One usage scenario is where a sender wants to maximize the economic return which may occur on the recipient reading the message. This scenario is relevant to advertisements, bulletins and newsletters. When there is no per-recipient cost, the optimal solution is to send to as many recipients as possible. This is where spamming originates. When a significant per-recipient cost is put in place, the optimal solution will be to select the recipients that are most likely to respond.

From an interpersonal communication viewpoint, the aim of Gridemail is to regulate communications in a way that is beneficial to both senders and receivers. Most of the benefit gained will have intangible properties such as goodwill or knowledge. We will not consider the monetary value of any payments as a benefit, as payments are used purely for regulation. Different combinations of benefits to the sender or receiver are illustrated in Figure 4. Messages can be placed anywhere in the graph.

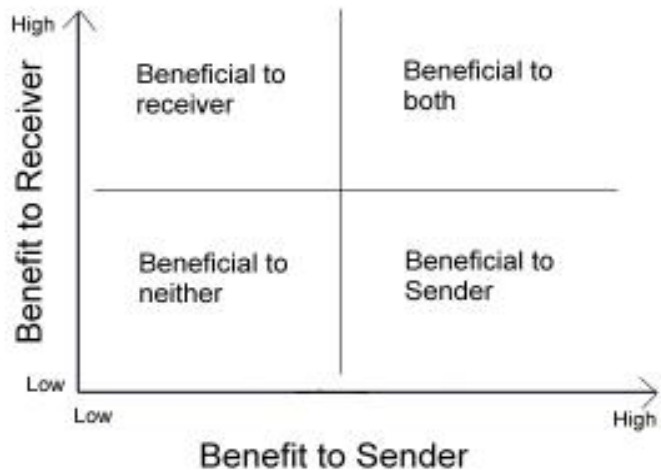

**Figure 4 – Classification grid for different types of messages.**

As each message arrives, the recipient's policy may accept, reject, or conditionally accept the message (based on payment of a charge). The set of messages waiting to be processed when the user logs in next time is a result of a sequence of decisions made according to the recipient's policy. We want to study possible policies which select messages that maximize the benefit to the recipient.

A variation of the above problem describes the job-shop decision model [49]. The objective is to find an optimal policy for accepting or rejecting jobs. The optimal policy will maximize the long-term profit by selecting jobs according to the job's characteristics. Jobs are processed immediately, hence accepting a job makes an immediate time commitment. In Gridemail, the decision to accept a message makes a future time commitment, and also reduces the time available to commit for subsequent messages.

Let us consider a limited time interval (0, x). During this time interval, a person has to choose between reading from k available messages and performing from j other tasks. The benefit and time taken for reading messages or doing tasks is illustrated in Figure 5. Each message or task is shown as a rectangle, where the horizontal dimension represents the time taken to process the message or perform the task. The area of the rectangle represents the net benefit. To make the problem interesting, the time interval (0, x) has to be too short to complete all tasks and read all messages with a net benefit.

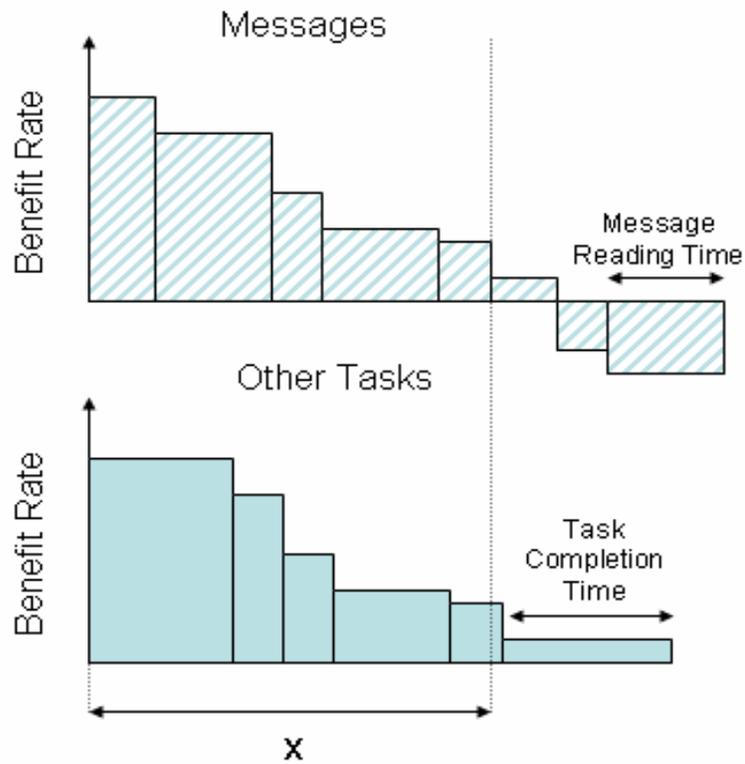

**Figure 5 – Illustration of benefit in reading mail or doing other tasks.**

The optimal solution in this scenario is to choose messages and tasks with the highest benefit rate sufficient to cover the whole interval, as shown in Figure 6.

If the arrival pattern of messages is known initially, and a probabilistic model is available for the benefits and times, decisions can be made following the principle of maximum expected utility [46, 47, 48]. The above solution assumes such information is available. However, that assumption can not be made initially in a real-life situation.

Probabilistic models of the benefits and processing times for messages can be built over time. These models will provide distributions for the benefit and processing time for a message as a function of the message properties such as sender, size and content, class of service, etc. Creating such models is discussed further in the next section.

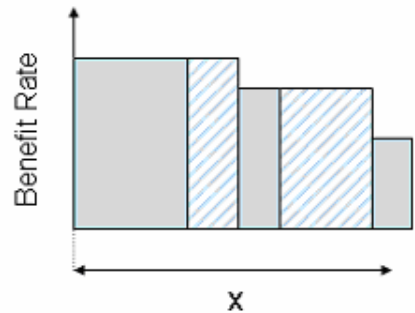

**Figure 6 – Illustration of maximizing total benefit.**

We will first study policies that are not pricing dependant. The simplest policy is to accept any message. We will use the following parameters for analysis:
(1) Messages arrive with exponential inter-arrival times
(2) The benefit and reading time of messages are independent and follow normal distributions. Mean time for reading is 3 minutes, and the mean benefit rate is 5 units per minute.
(3) The user logs in every 5 hours to check for messages, and reads all messages.
(4) The user has 15 minutes allocated exclusively for reading. If the time to read runs over that, she has to give up other tasks that have benefit rates of 10 units per minute.

If the arrival rate of messages is $\lambda$ per minute, the number of expected messages in 5 hours is $300\lambda$. The expected time taken to read messages will be $900\lambda$. As long as this time is less than 15 minutes (ie $\lambda \leq 1/60$), the expected net benefit will be $4500\lambda$. After the total reading time exceeds 15 minutes, the net benefit will be $4500\lambda - 10(900\lambda - 15)$.

Another simple policy is one which accepts messages until the expected time taken for reading messages is 15 minutes, and then rejects any subsequent messages. This policy will need a probabilistic model of how long each message will take to read. Like before, the expected net benefit will be $4500\lambda$ as long as $\lambda \leq 1/60$. Then it will remain at the same level as at $\lambda = 1/60$. The variation of expected net benefit with the arrival rate is shown in Figure 7.

A major problem with the above policies is that they would not inherently prevent spam or unimportant messages getting through. They might be acceptable when there is a favorable distribution of messages, but if there is a bias towards low value messages their performance will degrade.

Next we will consider policies that use some pricing controls. The simplest pricing policy is to have a fixed price for all messages. This policy is employed in micropayment systems that are currently available. Since they are effective in preventing spam messages getting through, they will be more robust than the two policies described before.

A policy based on pricing can regulate message acceptance in several ways. One way is to increase or decrease the price so that equilibrium is reached at a desirable level for the recipient. Limits on the expected processing time can optionally be used as in previous policies. Congestion pricing is slightly more complicated policy, where the price is increased with the number of messages already waiting to be processed.

Congestion pricing may be used without any limits on the expected processing time, as it will increasingly restrict acceptance rates.

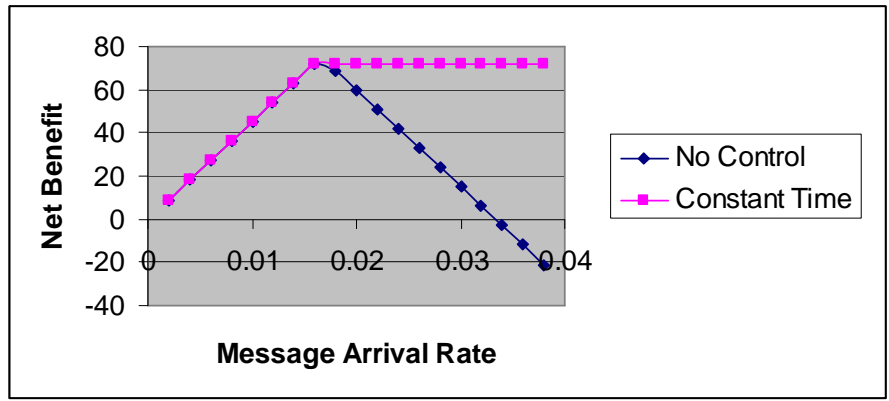

**Figure 7 – Variation of net benefit with arrival rate.**

The only safe assumption we can make about a sender is that he will be willing to pay more for message that benefit him. The recipient's benefit may not be a primary concern. If the only objective of the recipient is to maximize her benefit, a simple pricing policy is effective only when there is a direct relationship between the benefit to the sender and benefit to the recipient. If there is such a relationship, increasing the price can increase the expected benefit of accepted messages, as illustrated in Figure 8. Increasing the price (moving the vertical line to the right) will filter out messages that are less beneficial to the recipient and the sender.

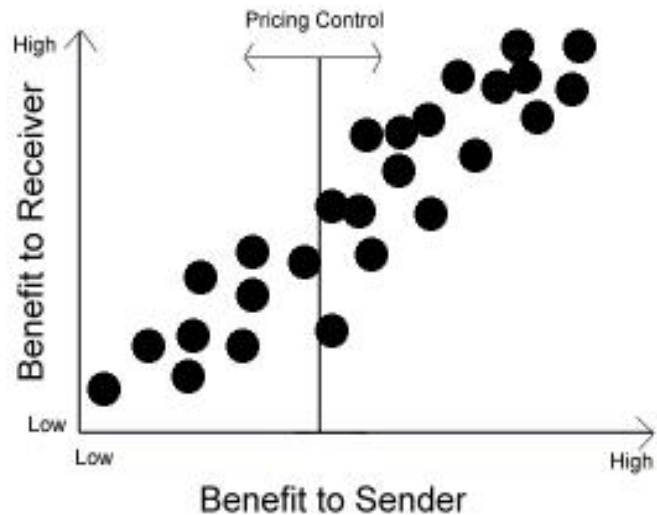

**Figure 8 – Effective pricing control.**

In reality there are messages that are beneficial mostly to the sender that recipients may want to or are obliged to read. An example would be a lecturer answering queries from students. There are other messages that are beneficial mostly to the recipient. Pricing policies may discourage the sender from sending such messages. The effect of pricing control on such relationships is shown in Figure 9. Increasing the price filters out messages that are beneficial only to the recipient.

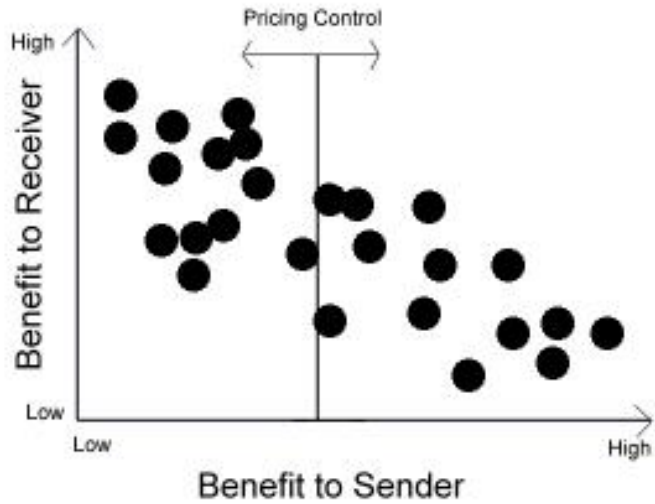

**Figure 9 – Ineffective pricing control.**

In Gridemail this problem is approached by having Classes of Service. Each COS can be used to regulate messages of a particular benefit characteristic. As illustrated in Figure 10, COS1 defines a class that delivers messages that are beneficial mostly to the receiver. Only trusted senders will be allowed to use this class of service. COS2 defines a class that delivers low benefit to the receiver. Pricing can effectively regulate messages in this class. COS3 represents another possible class that represents messages that are beneficial to the receiver and possibly to the sender. This class can represent useful communications between friends. A loose form of congestion pricing may be used to regulate messages in this class.

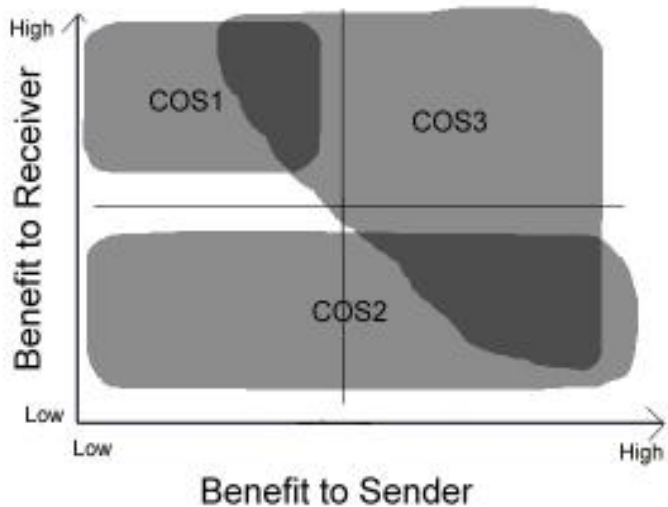

**Figure 10 – Classes of Service.**

## 9. Implementation Strategy for Gridemail

As described in the previous section, probabilistic models of the time taken to read messages are needed for some control policies. The time taken to process a message is roughly how long the message is viewed by the user, and can be measured by the browsing application. If there is a correlation between known properties of the message (such as the size and source) and the time taken, statistical analysis can be used to build a model.

There are various technical and social factors to consider when implementing Gridemail. From a technical perspective it would be ideal to start from a clean slate and develop all the infrastructure and protocols. From a social perspective, people and corporations already have a lot of practices and infrastructure in place already, and would be more likely to accept solutions that work with them.

The changes to implement Gridemail have to be minimized at the sender and recipient interfaces. An application similar to mail browsers can be used as the interface for composing and browsing messages. The Receiver service can also implement a message storage service, and the Post Office Protocol can be used largely unchanged to browse messages on the storage.

If changes are to be minimized, all infrastructure including SMTP servers can be used as is. However that will mean QoS selection and negotiation will have to be done manually by the sender. Payments and QoS selections can be specified in mail attachments created by accessing the Receiver Service through a web browser or a special application. These time consuming steps can be avoided by adding a QoS negotiation stage into SMTP so that it can act as the Sender Service or the Receiver Service as described in this paper. There will be no allowance for relaying, as messages are transferred directly from the Sender Service to Receiver Service.

The Payment and Security services can utilize existing infrastructure. We plan to use Gridbank [50] to implement the Payment Service.

The effect of QoS negotiation on the performance will be significant. This will have to be balanced against the reduction of unnecessary traffic.

## 10. Summary and Future Work

This paper presented an overview of Gridemail - a framework that overcomes problems in conventional email systems by regulating messages through economic methods. Policies based on pricing and quality of service has been suggested as effective control mechanisms.

We have discussed the concept of benefit of reading a message. However the policies discussed in this paper do not depend on a measurement or model of benefit. Such a model may possibly be built with feedback from the user and used as part of the policies.

Future work includes implementation of a prototype and development of various economic-based valuation strategies for handling different classes of messages. Their application in bulk and other types of messaging will also be studied.

[3] Judge, P.: Proposals according to the taxonomy of anti-spam systems, *ASRG website, http://www.irtf.org/asrg/survey_of_proposals.htm,* 2003.
[4] Asami, T. et al.: Taxonomy of SPAM and protection methods for Enterprise Networks, *ICOIN2002*, Berlin, pp 442-452, 2002.
[5] Naor, P.: Regulation of queue size by levying tolls. *Econometrica*, Vol 37, No 1, pp 15-24, 1969.
[6] Anbazhagan, M., Nagarajan, A.: Understanding quality of service for Web services. *IBM Developerworks website, http://www-106.ibm.com/developerworks/library/ws-quality.html*, 2002.
[7] Khong, W.: *The Law and Economics of Junk Emails*, Masters Thesis, University of Hamburg, 2001.
[8] Cobb, S.: The Economics of Spam, *EPrivacyGroup Website, http://www.eprivacygroup.com*, 2003.
[9] Cynthia, D. and Naor, P.: *Pricing via Processing or Combatting Junk Mail*, CRYPTO'92, pp 139-147, SpringerVerlag, 1992.
[10] Fakih, A.: Adaptive Mail Delivery Protocol, *IETF draft, http://www.ietf.org/internet-drafts/draft-fakih-amdp-00.txt,* 2003.
[11] Schaivone, V. et al: Trusted Email Open Standard – A Comprehensive Policy and Technology Proposal for Email Reform, *EPrivacyGroup Website, http://www.eprivacygroup.com*, 2003.
[12] Brussin, D.: SpamSquelcher – Reversing the economics of spam, *EPrivacyGroup Website, http://www.eprivacygroup.com*, 2003.
[13] Brightmail Spam Statistics, *http://www.brightmail.com/spamstats.html*.
[14] Coalition Against Unsolicited Bulk Email, Australia, http://www.caube.org.au
[15] Centipaid corporation, http://www.centipaid.com
[16] Heyman, D.P.: Optimal Operating Policies for M/G/1 queuing systems, *Operations Research*, Vol 16, pp 362-382, 1968.
[17] Lippman, S.A et al: Individual versus social optimization in exponential congestion systems, *Operations Research*, Vol 25, pp 233-247, 1977.
[18] Miller, B.: A queuing reward system with several customer classes, *Management Science*, Vol 16, pp234-245, 1969.
[19] Amaditz, Kenneth C.: Canning 'Spam' in Virginia: Model Legislation to Control Junk E-mail. *Vanderbilt Journal of Law & Technolog,* Vol 4, 1999.
[20] Bartels, David T.: Canning Spam: California Bans Unsolicited Commercial E-Mail, *McGeorge Law Review 30* , pp 420, 1999.
[21] Carroll, Michael W.: Garbage In: Emerging Media and Regulation of Unsolicited Commercial Solicitations, *Berkeley Technology Law Journal*, Vol.11, p233, 1996.
[22] Edwards, Lilian.: Canning the Spam: Is There a Case for Legal Control of Junk Electronic Mail?, *Law and the Internet: A Framework for Electronic Commerce*, ed. Lilian Edwards and Charlotte Waelde, 309–330. Oxford: Hart Publishing, 2000.
[23] Fasano,Christopher D.: Getting Rid of Spam: Addressing Spam in Courts and in Congress, *Syracuse Law & Technology Journal, Vol* 3, 2000.
[24] Gauthronet, S. and Etienne D.: Unsolicited Commercial Communications and Data Protection. *Brussels: Commission of the European Communities, Internal Market Directorate General, Contract no. ETD/99/B5-3000/E/96,* 2001.
[25] Hatchett, E.: The Spam Ban: The Feasibility of a Law to Limit Unwanted Electronic Mail, *Cyberspace Law Journal.* http://raven.cc.ukans.edu/~cybermom/CLJ/hatchett.html , 1998.
[26] Hawley, Anne E.: Taking Spam Out of Your Cyberspace Diet: Common Law Applied to Bulk Unsolicited Advertising via Electronic Mail, *University of Missouri at Kansas City Law Review 66*, pp 381–423, 1997.
[27] Kappel, Jennifer M.: Government Intervention on the Internet: Should the Federal Trade Commission Regulate Unsolicited E-Mail Advertising?, *Administrative Law Review 51*, pg 1011, 1999.
[28] Kosiba, Jeffrey L.: Legal Relief from Spam-Induced Internet Indigestion, *Dayton Law Review 25* , pg 187, 1999.
[29] Lee, Richard C.: Cyber Promotions, Inc. v. America Online, Inc, *Berkeley Technology Law Journal 13*, pg 417, 1999.
[30] Loren, Lydia Pallas: Regulating Cyberspace: A Case Study in SPAM, *http://www.cyberspacelaw.org/loren/index.html*.
[31] Miller, Gary: How to Can Spam: Legislating Unsolicited Commercial E-Mail, *Vanderbilt Journal of Entertainment Law & Practice 2, no. 1*, pg 127, 2000.
14